# END TO END SATELLITE SERVICING AND SPACE DEBRIS MANAGEMENT

Aman Chandra,[*] Himangshu Kalita,[†]
Roberto Furfaro,[‡] and Jekan Thangavelautham[§]

There is growing demand for satellite swarms and constellations for global positioning, remote sensing and relay communication in higher LEO orbits. This will result in many obsolete, damaged and abandoned satellites that will remain on-orbit beyond 25 years. These abandoned satellites and space debris maybe economically valuable orbital real-estate and resources that can be reused, repaired or upgraded for future use. Space traffic management is critical to repair damaged satellites, divert satellites into warehouse orbits and effectively deorbit satellites and space debris that are beyond repair and salvage. Current methods for on-orbit capture, servicing and repair require a large service satellite. However, by accessing abandoned satellites and space debris, there is an inherent heightened risk of damage to a servicing spacecraft. Sending multiple small-robots with each robot specialized in a specific task is a credible alternative, as the system is simple and cost-effective and where loss of one or more robots does not end the mission. In this work, we outline an end to end multirobot system to capture damaged and abandoned spacecraft for salvaging, repair and for deorbiting. We analyze the feasibility of sending multiple, decentralized robots that can work cooperatively to perform capture of the target satellite as a first step, followed by crawling onto damage satellites to perform detailed mapping. After obtaining a detailed map of the satellite, the robots will proceed to either repair and replace or dismantle components for salvage operations. Finally, the remaining components will be packaged with a de-orbit device for accelerated de-orbit.

## 1.0 INTRODUCTION

The growing space traffic expected due to rise in small satellites and CubeSat mega-constellations presents important opportunities and challenges. The increased access to space is expected to lower the cost of space services, including satellite communications, internet access and access to timely high-resolution earth imagery. However, if the growth remains unchecked and without effective space traffic management, we are likely to encounter uncontrolled collisions between high-value spacecraft and space debris. An effective space traffic management system needs to have in its toolbox technologies to identify, inspect, service/repair and set space debris in an accelerated deorbit path or parking orbit trajectory. The ideal technology to deorbit space debris needs to have a scalable architecture so that it can work for objects from 10 kg to 10 tons. Importantly, the technology needs to work with old and derelict spacecraft that were never intended to be accessed or serviced from space. These factors suggest the de-orbit device needs to work with satellites of various shapes and sizes and not rely upon standard latching/docking devices such those found on DARPA's Orbital Express or the PDGF (Power Data Grapple Fixture) found on the International Space Station (ISS).

---

[*] PhD Student, Aerospace and Mechanical Engineering, University of Arizona, 85721, USA.
[†] PhD Student, Aerospace and Mechanical Engineering, University of Arizona, 85721, USA.
[‡] Professor, Systems and Industrial Engineering, University of Arizona, 85721, USA.
[§] Assistant Professor, Aerospace and Mechanical Engineering, University of Arizona, 85721, USA.

In this paper, we present a novel solution to capturing and deorbiting old and derelict satellites. Our approach uses tethers to simplify capture and docking. The unique multi-functionality of tethers can simplify the object capture and deorbit task. We utilize a group of three 1U CubeSats that are interlinked by tethers. Each 1U CubeSat is fully autonomous, equipped with an onboard micro-gripper spine, 3D vision and navigation system, propulsion system and 3-axis attitude determination and control system. Importantly each 1U CubeSats is packaged with a toroidal inflatable de-orbit device. The inflatable utilizes solid sublimate that turns to gas in a vacuum to inflate a 1 m diameter toroidal cone.

Once space debris has been found, the three 1U CubeSats are deployed from a mothership/service vehicle. The 1Us separate with 0.5 m tether separating them into three lobes (star configuration) and approach the space debris. Upon contact, the modules wrap around the debris. Finally, the modules latch onto the debris using onboard micro-spine grippers. Once the modules are secured to the space debris, they begin a process of mapping. Through the mapping process, pieces of the satellite maybe identified for salvaging/repair/reuse. Alternately, the remaining space debris would be attached to inflatables that are deployed to accelerate the de-orbit process. For this proposed de-orbit device, the tethers maybe upgraded to be electrodynamic tethers that interact with the Earth's magnetic field to exploit Lorentz forces. Utilizing Lorentz forces, the debris is further propelled speeding up de-orbit.

In section 2, we compare the proposed deorbit solution with current state-of-the-art. In section 3, we present details of the de-orbit spacecraft concept to demonstrate the technology. In section 4, we analyze the dynamics of the tethered spacecraft modules and how they would grapple onto the space debris. In section 5 we analyze the propulsion requirements of the modules. In section 6 we analyze the de-orbit capability of the inflatable unit followed by conclusions and future work in section 7.

**2.0 BACKGROUND**

Deployable systems have been studied extensively as structures that facilitate atmospheric entry. Similar applications of such systems for aero-braking drag devices has received limited attention. Inflatable structures have seen development since the 1950's when NASA launched their ECHO satellite balloon program [1]. Inflatable technology received considerable attention for structural applications varying from gossamer sails, antennas, landing airbags and solar panels [2]. Ruggedized inflatables made of thermal fabrics started being researched for the challenging thermos-structural conditions during atmospheric re-entry. The first inflatable re-entry test was carried out in the year 2000 as a demonstration of inflatable re-entry and descent technology (IRDT) [3]. The structure consisted of an inflatable flexible heat shield and a parachute landing system. The experiment demonstrated significant improvement in payload to mass ratios due to highly efficient packing and low-weight of the inflatable. Structural and thermal performance was observed to be enough to survive atmospheric-entry into Earth.

The success of the IRDT established inflatable technology as a robust and low-cost alternative to existing re-entry technologies. A second successful experiment was seen when NASA launched the inflatable re-entry vehicle experiment (IRVE) in 2006. The IRVE is a 3-meter diameter, 60° half angle sphere cone consisting of an inflatable aero-shell structure. The purpose of the experiment was to validate aero-shell performance for atmospheric re-entry [4]. For a braking or de-orbit device, the encountered thermo-structural loads are an order of magnitude lower than atmospheric re-entry. This simplifies the design and allows compact and lightweight gossamers to be used. Andrews Space has developed a prototype that has undergone ground based tests as an inflatable nanosat de-orbit and recovery system for CubeSat payloads [5]. Figure 1 shows an illustration of their design. Italian concept IRENE [6] is undergoing tests with a spherical cone designs but is intended for much larger payloads.

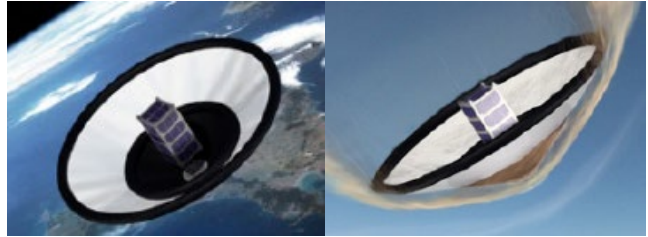

**Figure 1**. Proposed d2U CubeSat DRS mechanism [10]

Carandente et al. [7] present concepts for aero-braking structures from a de-orbit and re-entry point of view. The concepts presented are ones that offer substantial reduction of ballistic coefficient of these structures. The authors point out that a lower ballistic coefficient leads to a reduction in peak mechanical and thermal loads experienced by the structure while achieving higher deceleration rates. This highlights the potential of using large inflatable gossamers that can be packaged into very small volumes for nano-satellite payloads.

Among large scale gossamer structures, Global Aerospace Corporation proposed the Gossamer Orbit Lowering Device (GOLD) to de-orbit spent stages and old or derelict satellites [8]. The concept consists of deploying a large inflatable sphere several meters in diameter that offers exceptionally low ballistic coefficients in Lower Earth Orbits. Fig. 4 shows the concept. Gossamer sails made of Kapton have also been studied in considerable detail [9]. While sails potentially offer more efficient packing ratios than inflatables, their structural reliability for aerobraking is not clearly established. While encountering loads due to atmospheric drag, a pneumatic pressure system has been used to provide necessary resistive stiffness. In the case of sails, additional structural re-enforcement is needed which has reduces packing efficiency and increases deployment complexity.

Pneumatic inflatables have shown robust structural behavior while maintaining ease of scaling into large sizes. The focus of our research is on inflatable structures. Pneumatic inflatable require a gas source. This can be in the form of a compressed gas or gas producing chemical reaction. Inflatables using solid state sublimates as gas sources have shown promising results for Low Earth Orbit operation [10]–[12].

Tethering of a spacecraft is not a new concept and has been present since the 1960's with Gemini VI and VII tether experiments [14, 18]. This was followed by more in-depth experiments during the space shuttle era in the 1980s and 1990s. This included Charge-1 in 1983, Charge-2 in 1985, OEDIPUS-A (Observations of Electric-Field Distribution in the Ionospheric Plasma—a Unique Strategy) in 1989 [14] and OEDIPUS-C in 1995 [14], TSS-1 in 1992 and TSS-1R in 1996 [14], SEDS-1 (Small Expendable Deployer System) in 1993 and SEDS-2 in 1994 [14], PMG (Plasma Motor Generator) in 1993 [14], TiPS (Tether Physics and Survivability) experiments in 1996 [14], YES (Young Engineers Satellite) in 1997 [16], YES2 in 2007 [17], ATEx (Advanced Tether Experiment) in 1998 [6], MAST (Multi-Application Survivable Tether) in 2007 [20].

These experiments tested tether deployment, attitude control stabilization, gravity-gradient stabilization, power generation, drag-generation for de-orbiting and boosting/rebooting to higher orbits [14-15,18]. Tethered experiments have also been used to generate artificial gravity, facilitate payload rendezvous and capture and have been shown to enable aero-assisted maneuvering. Our earlier work has also utilized multiple tethered robots to hop and climb rugged surfaces [22, 24]. In addition, we have developed neural-network control systems that can autonomously traverse maze-like environments that would be expected on a satellite to determine mounting locations or locations to remove existing components or install de-orbit devices [29].

The versatility of using tethers for a wide variety of applications make them an important technology for both debris capture but also for orbital-traffic management. For all of these reasons, there is an important advantage to grapple two free floating masses in space and attach tethers to them. In the following section, we will describe the tether dynamics and ways for using robots to attach tether between free flying objects in Earth orbit.

**3.0 SYSTEMS DESIGN**

The proposed solution consists of three 1U CubeSats modules. Each module is deployed from a standard 3U P-POD deployer. One or more 3U P-POD deployer can be easily be mounted to a service spacecraft. Each 1U CubeSat is 1.3 kg and has 10 cm × 10 cm × 10 cm dimensions (Figure 2 and 3). The 1U CubeSat has a gripper surface that can latch on rough and metallic surfaces. Inside, there is an attitude determination and control system containing 3-axis reaction wheels and magnetorquers. The modules are powered using onboard 38 Whr lithium ion batteries and body-mounted solar panels providing up to 2 W. The onboard computer board combines a UHF transceiver, with dual-core ARM processor and an FPGA subsystem for 3D vision and navigation. Each module also has a sublimate-based propulsion system with a total delta-v of 0.4 m/s. Each module also contains a 1-m inflatable spherical deorbit device.

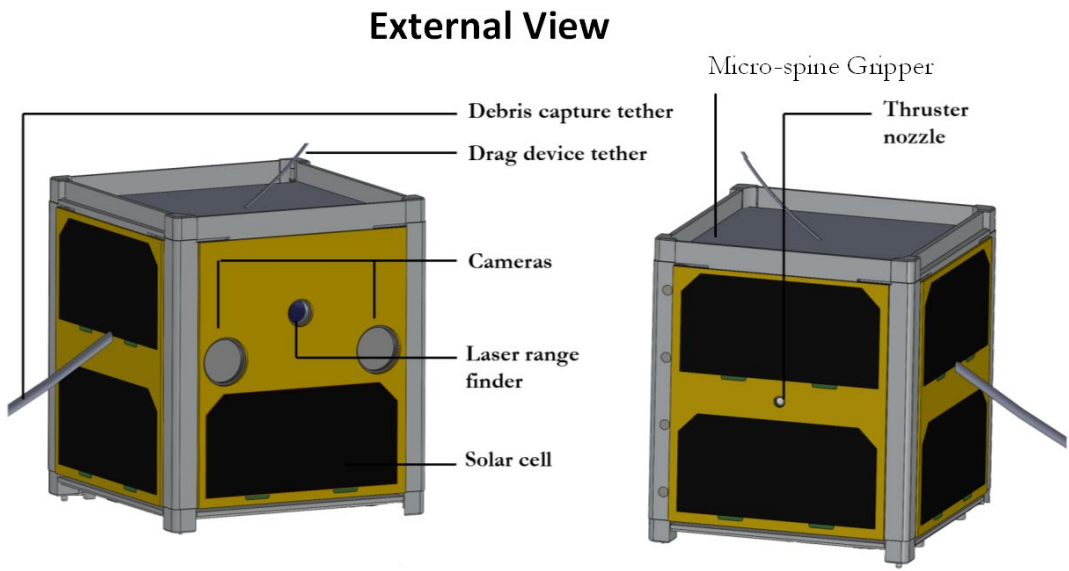

**Figure 2**. External View of 1U CubeSat De-Orbit Module

The de-orbit device inflates using a sublimate such as benzoic acid that turns into a gas under a vacuum. The inflatable can withstand small punctures and keep being inflated or a few months to a year using a few 10s of grams of sublimate. Each module is interlinked by a multi-functional tether. The tether serves multiple purpose by being able to snag or even wrap onto orbital debris. In addition, a tether is attached from the body of the module to the de-orbit device. If the tether were to be electrodynamic, it can exploit Lorentz force to accelerate a deorbit or parking maneuver. Each CubeSat operates autonomously and is interlinked by tethers to form three lobes. Alternately, they may also be teleoperated when performing high-risk maneuvers. The entire system is expected to be low-cost, disposable and can be readily be mass produced once validated in space.

Figure 4 shows an artistic view of the de-orbit system deployed on Maxar's Worldview 4 which stopped functioning in Dec. 2018. Once a service spacecraft rendezvous with the space debris, it will deploy a set of three 1U CubeSats interlinked by tethers. The modules will intercept the debris in a star configuration followed by snagging and wrapping around the debris. Once the modules are in close proximity with the debris, they will latch on the surface using the micro-grippers. Using the micro-grippers the spacecraft may attempt to reposition themselves on the debris. Once in the right position, each spacecraft will deploy the drag-device, namely the inflatable and maintain a tracking beacon.

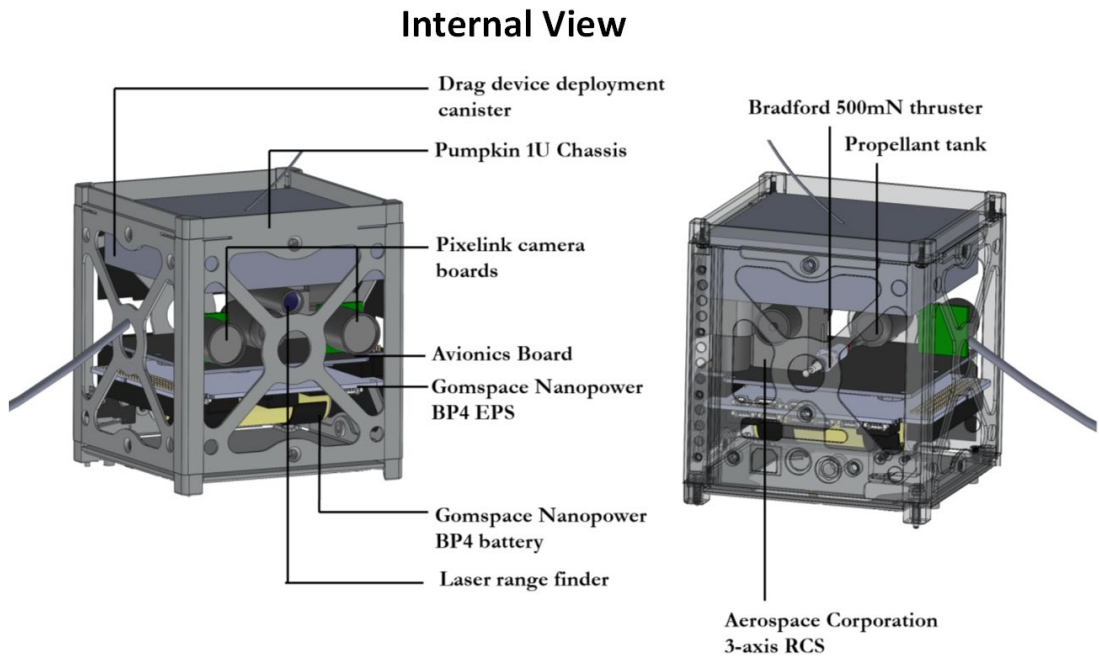

**Figure 3.** Internal View of 1U CubeSat De-Orbit Module

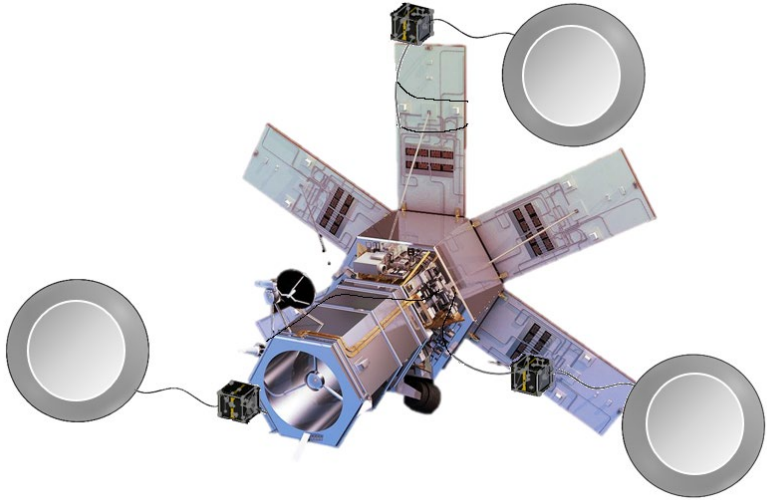

**Figure 4.** Artistic Image of De-Orbit System Deployed on Maxar's Worldview 4.

## 4.0 TETHERED CAPTURE DYNAMICS

Our approach to using tethered CubeSats to capture old and derelict satellites comes from earlier designs of multiple spherical hopping robots that would be interlinked by tethers to hop and climb low-gravity rugged terrain [21-22, 24-26]. The tethers connecting the CubeSats can be most efficiently described as a flexible body consisting of a series of point masses connected by massless springs and dampers in parallel. Using the Kelvin-Voigt model, the tether can be modeled as a viscoelastic material having the properties both of elasticity and viscosity through a combination of spring-dampers resulting in different tension laws. Tension on a rope element linking the $i^{th}$ node to the $j^{th}$ node can be expressed as Eq. (1).

$$T_{ij} = \begin{cases} [-k_{ij}(|r_{ij}| - l_0) - d_{ij}(v_{ij} \cdot \hat{r}_{ij})]\hat{r}_{ij} & if\ |r_{ij}| > l_0 \\ 0 & if\ |r_{ij}| \leq l_0 \end{cases} \quad (1)$$

where, $k_{ij}$ is the stiffness parameter of the tether element $ij$ which depends on the material properties and geometry of the tether, $d_{ij}$ is the damping coefficient of the tether element $ij$. $r_{ij}$ and $v_{ij}$ are the relative position and velocity between the $i^{th}$ node and the $j^{th}$ node. $\hat{r}_{ij}$ is the normalized unit vector along the position vector. Also, $l_0$ is the nominal un-stretched length of the tether element. During the wrapping and docking phase, multiple contact events will occur between the tether and the target satellite and also among different part of the tethered system. As a result, efficient collision detection and accurate representation of contact dynamics becomes key to the fidelity of the simulation to reality. The target spacecraft is modeled as a convex polyhedral and the Gilbert, Johnson and Keerthi (GJK) collision detection algorithm is used to detect collision between the tether and the target satellite and also to calculate the penetration depth during every collision.

After detecting the collision, Hertz contact force model has been implemented to model the contact dynamics. When two bodies collide, local deformations occur resulting in penetration into each other's space. The penetration results in a pair of resistive contact forces acting on the two bodies in opposite directions. Every collision consists of a compression phase and a restitution phase which can be modeled as a non-linear spring-damper as shown in Eq. (2).

$$f_N = K\delta^n + d_c\dot{\delta} \qquad (2)$$

where, $K$ is the stiffness parameter, which depends on the material properties and the local geometry of the contacting bodies, $\delta$ is the penetration depth, $d_c$ is the damping coefficient, $\dot{\delta}$ is the relative velocity of the contact points, projected on an axis normal to the contact surfaces and $n = 3/2$. Each collision between the tether and the target satellite results in a tangential frictional component of contact force which is computed using Coulomb's law of dry friction which opposes the relative motion. It has been experimentally found that the transition of friction force from zero to nonzero relative velocity is not instantaneous, but it takes place during a short period of time. This transition called the Stribeck effect is implemented to the equations of motion of the multibody system using the Anderson function to avoid stiction as shown in Eq. (3).

$$f_t = f_N \left( \mu_d + (\mu_s - \mu_d) e^{-\left(\frac{v_{i,j}}{v_s}\right)^p} \right) \tanh(k_t v_{i,j}) \qquad (3)$$

where, $\mu_s$ is the coefficient of static friction, $\mu_d$ is the coefficient of dynamic friction, $v_{ij} = v_i - v_j$ is the relative speed, $v_s$ is the coefficient of sliding speed that changes the shape of the decay in the Stribeck region, exponent $p$ affects the drop from static to dynamic friction and the parameter $k_t$ adjusts the slope of the curve from zero relative speed to the maximum static friction. To compute the aerodynamic forces acting on the tether, the model presented by Aslanov and Ledkov is implemented. One of the fundamental assumptions of the model is that every half of the tether part connecting two-point masses is considered rigid and hence moves at the same speed of the node. The aerodynamic force acting on a node $i$ can then be computed as shown in Eq. (4).

$$F_{ai} = \frac{\rho v_i d}{4} c_d \left( \frac{n_i}{r_{i,i-1}} + \frac{n_{i+1}}{r_{i+1,i}} \right) \qquad (4)$$

where, $\rho$ is the atmospheric density, $v_i$ is the velocity of node $i$, $d$ is the tether diameter, $c_d$ is the drag coefficient, $r_{i,i-1}$ is the distance between node $i$ and $i-1$, and $n_i = (v_i \times r_i) \times r_i$.

The block diagram to simulate the docking mechanism for the tethered system is shown in Figure 5. The algorithm first computes the elastic and damping tension forces along with the aerodynamic forces acting on each node and then integrates the dynamic equations of motion to compute its positions and velocities. The collision detection algorithm is then carried out to detect impending collisions. The colliding nodes along with their penetration depth and relative velocities are computed and the corresponding contact normal and tangential forces are calculated which are then used to integrate the dynamic equations of motion [21, 22, 24].

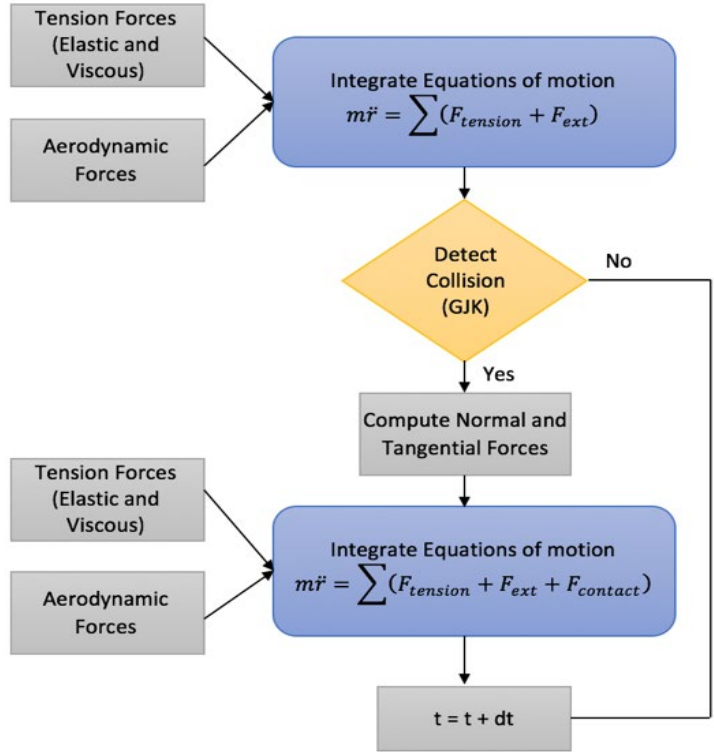

**Figure 5.** Simulation architecture for tether systems used.

To fully analyze the dynamics of the tethered system and the contact model, simulations are performed on a simplified cubical target satellite. The tethered system is modeled using 121 nodes, connected to three 1U CubeSats (Figure 6). The tethered system is deployed in a 'star' configuration with initial relative velocity w.r.t the target satellite of 15 m/s along the y-axis. The tether material modelled is Technora used to suspend the NASA Mars rover Opportunity from its parachute during descent.

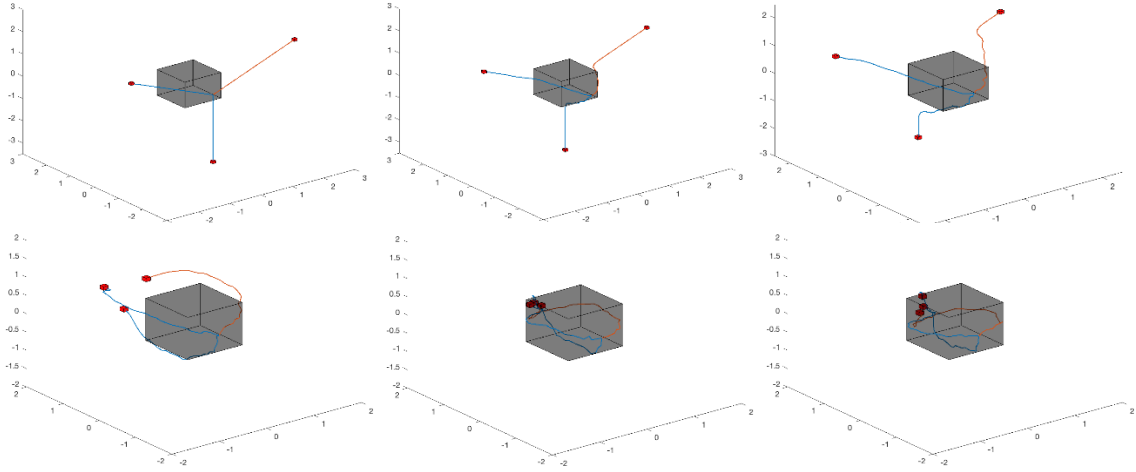

**Figure 6.** Capture process using three 1U CubeSats at different time step with no rotation of target satellite.

For our simulation the Young's modulus of the tether is 25 GPa, the damping ratio is set to 0.3 and the density at 1390 kg/m³. For the contact dynamics, the stiffness parameter is considered as 500 N/m and the damping coefficient as 0.5. For the friction model, the coefficient of static and dynamic friction is considered 0.7 and 0.5 respectively and the parameters as $v_s = 0.001$, $p = 2$, and $k_t = 10000$. The dimension of the target satellite is $1.15 \times 1.15 \times 1.15 \ m$. Figure 6 shows the capture and docking process at different time step. Further simulations were performed with the target satellite rotating with a constant angular velocity of [1 0.5 0.2] deg/s as shown in Figure 7. It can be seen that the tethered robotic system was able to capture the target satellite.

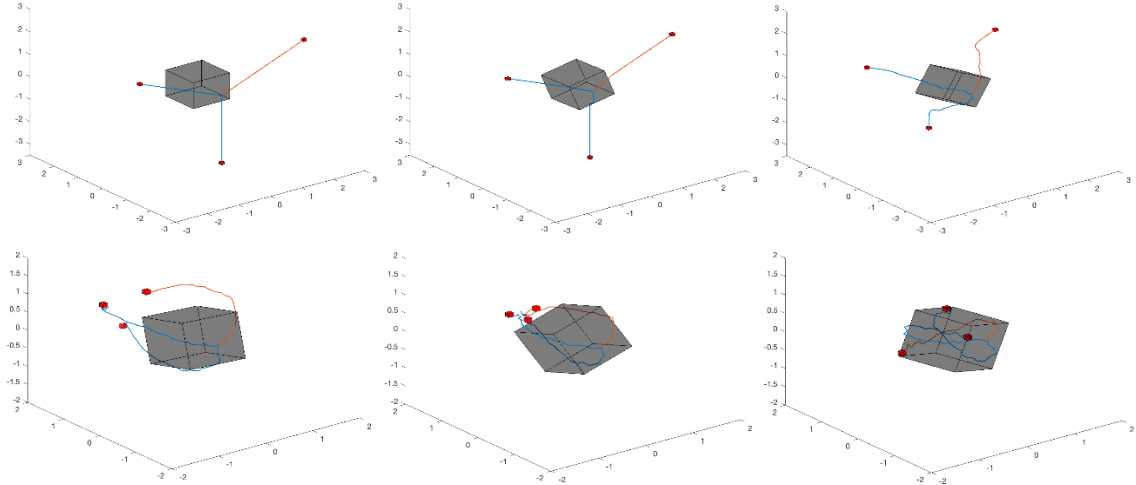

**Figure 7.** Capture and Docking process at different time step with target satellite rotating with a constant angular velocity of [1, 0.5, 0.2] deg/s.

### 5.0 PROPULSION

The purpose of the propulsion system is to align each CubeSat into the proper star configuration with the tethers in tension. With 0.2 kg allocated to the entire propulsion system, there is a reliance on microelectromechanical systems (MEMS) to decrease volume and mass. In this work, we baseline a sublimate-based cold-gas propulsion system, with four discretized nozzles on a MEMS chip [23]. Each nozzle can be individually actuated to provide a small attitude correction, or all four are fired for a thrust of 0.5 Newtons (Figure 8). Alternate options include water electrolyzed into hydrogen and oxygen on-demand for propulsion [27], in addition to water-steam [28]. However, both options are too complex to integrate into a 1U CubeSat at the present time.

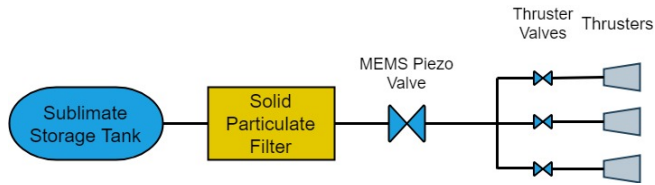

**Figure 8.** Propulsion System Schematic

To put each tether in tension, the nozzle chip is placed on the tethered face, such that the propulsive force acts away from the central tether connection. Each nozzle is located so as to not interfere with the tether. Sublimates are used for their low chamber pressure, reducing design complexity and allocating more mass to the propellant for a higher delta-v. Sublimation is an endothermic process, requiring heat addition often by resistive heating for molecules to escape attractive forces. As the temperature of the system increases, the vapor pressure increases. This leads to one of the large advantages of using sublimating propellant as opposed to gas storage: the sublimate acts as its own pressure regulator through temperature control. If a traditional cold gas system is used, the storage tank needs to be more robust for the higher storage pressure and the thrust monotonically decreases with each burn.

The structural mass of the propulsion system is 140 grams, leaving 60 grams of propellant. Utilizing the Tsiokolsky rocket equation, Equation (5), is used to estimate the total delta-v capable for each CubeSat:

$$\Delta v = g_0 I_{sp} ln\left(\frac{m_0}{m_f}\right) \quad (5)$$

Where $g_0$ is Earth gravity, $I_{sp}$ is he measure of efficiency of the thruster, $m_0$ is the full mass of the system, and $m_f$ is the final mass after all propellant has been depleted. Thus, each CubeSat is capable of 0.36 m/s of delta-v with a thruster $I_{sp}$ of 0.6. Propulsion also allows for minor corrections to the cluster's trajectory. This ensures the target can be adequately tethered by arranging the CubeSats in the optimal position.

## 6.0 CUBESAT DRAG DEVICE DESIGN AND ANALYSIS

De-orbit performance is a function of drag force experienced by the device. As the spacecraft altitude increases, the surrounding atmosphere continues to rarify thereby reducing drag. The effect of height on atmospheric drag is described by (6)

$$\rho \approx \rho_o e^{-\Delta h/h_o(h)} \quad (6)$$

The atmospheric density $\rho$ at a given altitude and $\rho_o$ at a second altitude with difference in height of $\Delta h$ are related exponentially as shown. $h_o(h)$ termed as scale height is a function of altitude. We begin by studying the nature of forces encountered towards two major applications. For a circular orbit at altitude $H$ above the Earth, the average change in acceleration due to drag is as shown in (7):

$$\Delta a_{rev} = -\frac{2\pi\rho a^2}{b_c} \tag{7}$$

To enable atmospheric burn up at 100 km altitude, requires the ballistic coefficient be a function of altitude and can be written as:

$$b_c = \frac{(R_e + H)^2 (2R_e + H + 10^5)(H - 10^5)}{2\pi\mu\rho(R_e + 10^5)^2} \tag{8}$$

Figure 9 shows basic structural design elements for such a system on board a CubeSat

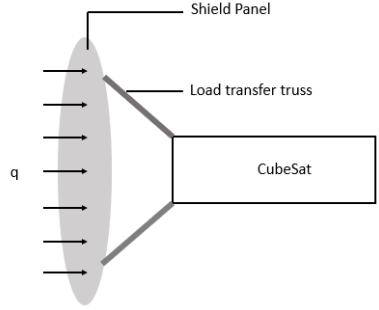

**Figure 9**. Structural design elements of the de-orbit device.

The pressure distribution on the surface of the shield can be assumed to be uniform in nature and is calculated from the aerobrake inertia force due to a constant rate of deceleration. The pressure $p$ is given as shown in (9) as:

$$p = \frac{M_s a}{g A_{ab}} \tag{9}$$

Here $M_s$ is the mass of the spacecraft attached to the braking system, $a$ is the deceleration rate, $g$ is a constant defined by Newton's Law as Force = (mass × acceleration). $A_{ab}$ is the area of the aerobraking structure. We extend their methodology to include design of inflatable membrane structural units. Based upon structural function, the aero-braking device consists of a shield and support structure. The fundamental structural sizing equation is a shown below.

$$w_{max} = \alpha \frac{qA^4}{D_{HP}} \tag{10}$$

Here $w_{max}$ represents a bound on maximum mass of the structure for achieving a bending stiffness $D_{HP}$ for encountered drag force $q$ over area $A$. Based on sizing requirements, we propose a concept device [13] as shown in Figure 10.

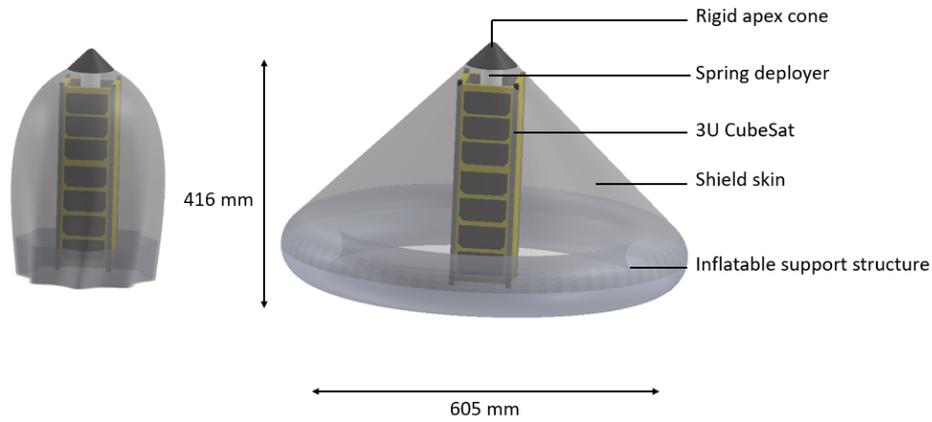

**Figure 10**. Toroidal inflatable aero-shell [13].

Aero-braking or de-orbit performance of the inflatable is characterized by estimating its drag co-efficient. Computed drag coefficient values were used to calculated obtained deceleration using

$$\Delta P_{rev} = -6\pi^2 \left(C_D A / m\right) \rho a^2 / V \tag{11}$$

Here $\Delta P$ represents a change in orbital period for a circular orbit characterized by the CubeSats velocity V and mass m. The calculated loads are compared with expected structural behavior to understand their ability to maintain structural integrity. Table 1 shows estimated drag coefficients, ballistic coefficients and estimated orbit decay for a 3U CubeSat using both inflatable concepts with an estimated total mass of 4 kg.

**Table 1.** Comparison of braking concepts [13]

| Design Concept | Mass (kg) | Surface Area (m²) | Drag Coefficient | Ballistic Coefficient (kg/m²) |
|---|---|---|---|---|
| 3U CubeSat | 3.5 | 0.01 | 2 | 175 |
| Concept Design | 4 | 0.346 | 2.7 | 4.28 |

The above table shows a dramatic decrease in ballistic co-efficient upon adding the inflatable structures onto the 3U CubeSat. This is due to much larger surface areas at very low additional mass. A larger drag coefficient is possible in thanks to an optimized spherical cone geometry. Based on calculated co-efficient values, we go on to calculate estimated de-orbit lifetimes for each case. Table 2 shows estimated orbit decay times from various altitudes of a circular orbit. The reduction in de-orbit lifetimes is in agreement with much lower ballistic co-efficient.

**Table 2.** Comparison of expected de-orbit times [13]

| Altitude (km) | Disposal Life-time (years) | |
|---|---|---|
| | 3U CubeSat | Concept |
| 400 | 1.2 | 0.045 |
| 500 | 6.3 | 0.28 |
| 600 | 23.5 | 1.5 |
| 700 | >25 | 6.2 |
| 800 | >25 | 17.8 |

## 7.0 CONCLUSIONS

A critical element of space-traffic management is the ability to move old and derelict satellites into safe-parking or disposal orbits to prevent future collisions with high-value satellite assets. In this paper we presented a scalable, low-cost de-orbit system consisting of multiple autonomous 1U CubeSats interlinked by tethers to capture space debris and deploy inflatable to accelerate the de-orbit process. The proposed approach can work for satellites never intended to be serviced or repaired in space. Using electrodynamic tethers, the de-orbiting process could be further accelerated using Lorentz forces. Plans are underway to validate the component technologies in the laboratory and raise them to TRL-5. Beyond these laboratory experiments, we intend to demonstrate the technology in Low Earth Orbit.